\documentclass[aps,prb,10pt,twocolumn]{revtex4-1}

\usepackage{graphicx,bm,amsmath,overpic}

\newcommand{\bfr}{{\bf r}}
\newcommand{\bfB}{{\bf B}}
\newcommand{\bfm}{{\bf m}}
\newcommand{\bfc}{{\bf c}}

\newcommand{\ua}{\uparrow}
\newcommand{\da}{\downarrow}

\begin{document}

\title{Exchange-correlation magnetic fields in spin-density-functional theory}

\author{Edward A. Pluhar III}
\author{Carsten A. Ullrich}
\affiliation{Department of Physics and Astronomy, University of Missouri, Columbia, Missouri 65211, USA}

\date{\today }

\begin{abstract}
In spin-density-functional theory for noncollinear magnetic materials, the Kohn-Sham system features exchange-correlation (xc) scalar potentials and magnetic fields.
The significance of the xc magnetic fields is not very well explored; in particular,
they can give rise to local torques on the magnetization, which are absent in standard local and semilocal approximations.
We obtain exact benchmark solutions for two electrons on four-site extended Hubbard lattices over a wide range of interaction strengths, and compare exact xc potentials
and magnetic fields with approximations obtained from orbital-dependent xc functionals. The xc magnetic fields turn out to play an increasingly important role as
systems becomes more and more correlated and the electrons begin to localize; the effects of the xc torques, however, remain relatively minor.
The approximate xc functionals perform overall quite well, but tend to favor symmetry-broken solutions for strong interactions.
\end{abstract}

\maketitle

\section{Introduction}

Spin-density-functional theory (SDFT) \cite{Barth1972,Gunnarsson1976,Gidopoulos2007} is an in principle exact framework for systems of $N$
interacting electrons subject to the many-body Hamiltonian
\begin{equation} \label{1}
\hat H = \sum_{j}^N\left[-\frac{\nabla_j^2}{2} + V(\bfr_j) + \bm{\sigma}_j\cdot \bfB(\bfr_j)  \right]
+ \frac{1}{2}\sum_{j\ne k}^N \frac{1}{|\bfr_j - \bfr_k|} \:.
\end{equation}
Here, $\bm{\sigma}_j$ is the vector of Pauli matrices acting on the spin of the $j$th electron,
$V(\bfr)$ is an external scalar potential, and $\bfB(\bfr)$ is an external magnetic field which
can be noncollinear, i.e., whose magnitude and direction can vary in space.
The Bohr magneton, $\mu_B = e \hbar/2m$, has been absorbed in the definition of the magnetic field,
and we use atomic units ($e = m = \hbar = 4\pi \epsilon_0 = 1$) throughout.

The corresponding noninteracting system in SDFT obeys the Kohn-Sham equation
\begin{equation} \label{2}
\left[\left(-\frac{\nabla^2}{2} + V_{\rm KS}(\bfr) \right)I + \bm{\sigma} \cdot \bfB_{\rm KS}(\bfr) \right]
\Psi_i(\bfr) = \epsilon_i \Psi_i(\bfr) \:,
\end{equation}
where $I$ is the $2\times 2$ unit matrix and the Kohn-Sham orbitals $\Psi_i(\bfr)$ are two-component spinors.
The scalar Kohn-Sham potential is the sum of external, Hartree, and exchange-correlation (xc) potentials,
$V_{\rm KS} = V + V_{\rm H} + V_{\rm xc}$, and the Kohn-Sham magnetic field contains external and xc contributions,
$\bfB_{\rm KS} = \bfB + \bfB_{\rm xc}$. In general, $\bfB_{\rm xc}$ can be nonvanishing even if $\bfB=0$,
i.e., for open-shell or spontaneously magnetized systems.

The effects of $\bfB_{\rm xc}$ are present in a wide variety of systems, including materials  with noncollinear magnetism,
canted or spiral magnetization, or frustrated spins. \cite{Freeman2004,Goings2018,Trushin2018} In these examples,
spin-orbit coupling often plays an important role. The SDFT formalism can be generalized to include spin-orbit coupling; \cite{Trushin2018,Krieger2015}
in this paper we ignore spin-orbit coupling to keep the discussion simple.

Formally, $V_{\rm xc}$ and $\bfB_{\rm xc}$ are defined as functionals of the density
$n(\bfr) = \sum_i^N \mbox{tr}[\Psi_i(\bfr)\Psi_i^\dagger(\bfr)]$ and the magnetization \cite{footnote1}
$\bfm(\bfr) = \sum_i^N \mbox{tr}[ \bm{\sigma} \Psi_i(\bfr)\Psi_i^\dagger(\bfr)]$;
in practice, $V_{\rm xc}$ and $\bfB_{\rm xc}$  need to be approximated. Almost all approximations used in SDFT for noncollinear magnetism
are based on common local and semilocal xc functionals such as the local spin-density approximation, LSDA, or the generalized
gradient approximations, GGAs. The standard implementation of LSDA or GGAs assumes a local spin quantization axis which is aligned with the
local $\bfm(\bfr)$; \cite{Kubler1988,Sandratskii1998} this produces a $\bfB_{\rm xc}(\bfr)$ that is parallel to $\bfm(\bfr)$
everywhere.

In general, $\bfB_{\rm xc}(\bfr)$ can have components perpendicular to $\bfm(\bfr)$, which give rise to local xc torques,
\begin{equation} \label{tau}
\bm{\tau}_{\rm xc}(\bfr) = \bfm(\bfr) \times \bfB_{\rm xc}(\bfr) \:.
\end{equation}
The zero-torque theorem \cite{Capelle2001} mandates that
\begin{equation} \label{zerotorque}
\int d\bfr \:\bm{\tau} _{\rm xc}(\bfr)=0 \:,
\end{equation}
since a system cannot exert a net torque on itself.
For the LSDA and GGAs, the local xc torque is zero everywhere, so the zero-torque theorem is trivially satisfied.
As we will see, other approximations may not satisfy the zero-torque theorem a priori, but there are ways in which it can be enforced.

Several approximations for $\bfB_{\rm xc}$ that include xc torque effects have been proposed in the literature,
based on the spin-spiral state of the electron gas, \cite{Kleinman1999,Eich2013a,Eich2013b,Pittalis2017}
modified GGAs, \cite{Katsnelson2003,Peralta2007,Scalmani2012,Bulik2013} the exchange-only optimized effective potential (OEP), \cite{Sharma2007}
or a source-free construction. \cite{Sharma2018} These methods were applied to a variety of noncollinear spin
systems; however, a rigorous assessment has been hampered by the absence of exact or at least highly accurate reference calculations.

In this paper, we perform benchmark studies for a simple model system, namely, two electrons on a linear 4-point extended Hubbard lattice
($1/4$ filled Hubbard tetramer) with noncollinear magnetic fields. In Section \ref{sec:II} we will give the details of the model, and explain why it is an appropriate
choice for studying xc torques.

Results will be presented for two scenarios: a lattice Hamiltonian without any symmetries (Sec. \ref{sec:III}), and a Hamiltonian with
$C_2$ symmetry (Sec. \ref{sec:IV}). In both cases, we obtain the exact ground-state energy, densities, and magnetizations from the exact solutions,
and compare with approximate Kohn-Sham results. \cite{Ullrich2018,Ullrich2019} We reverse-engineer the exact xc potentials and magnetic fields
for interaction strengths from the weakly to the strongly correlated regime, and discuss how the Kohn-Sham system mimics the behavior of the
many-body system as the electrons begin to localize. We summarize our results in Section \ref{sec:V} and conclude with a general outlook.

\section{SDFT for the extended Hubbard model with noncollinear magnetic fields}\label{sec:II}

We consider $P$-point Hubbard chains with on-site (o.s.) and nearest-neighbor (n.n.) interactions
and with external scalar potentials and magnetic fields, $V_k$ and $\bfB_k$,
where $k$ denotes the lattice sites. The continuum Hamiltonian (\ref{1}) is thus replaced by the extended Hubbard model
Hamiltonian \cite{Hirsch1984,Zhang1989,Laad1991,Schumann2010}
\begin{eqnarray}\label{Hubbard}
\lefteqn{
\hat H_{\rm lat}= -t\sum_{\langle k,l\rangle \sigma} (\hat c^\dagger_{k\sigma}\hat c_{l\sigma} + \hat c^\dagger_{l\sigma}\hat c_{k\sigma}) + U_0 \sum_{k}
\hat c^\dagger_{k\ua}\hat c_{k\ua}\hat c^\dagger_{k\da}\hat c_{k\da} } \nonumber\\
&+&
U_1 \sum_{\langle k,l\rangle } (\hat \bfc^\dagger_{k}\hat \bfc_{k})(\hat \bfc^\dagger_{l}\hat \bfc_{l})
+ \sum_{k}\left[ V_k \hat\bfc^\dagger_{k}\hat \bfc_{k} + \bfB_k \cdot \hat\bfc^\dagger_{k} \bm{\sigma} \hat \bfc_{k} \right] \!.
\end{eqnarray}
Here, $\hat c_{k\sigma}^\dagger$ and $\hat c_{k\sigma}$ are creation and annihilation operators
for electrons with spin $\sigma=\ua,\da$ on site $k$, and $\hat\bfc_{k}^\dagger$ and $\hat\bfc_{k}$ are these operators arranged as two-component row and
column vectors. $\langle k,l\rangle$ denotes nearest-neighbor lattice sites. We fix the hopping parameter as $t=0.5$.
The o.s. interaction strength $U_0$ will be varied, and the n.n. interaction strength will be taken as $U_1 = U_0/2$.

We obtain the two-electron ground state of $\hat H_{\rm lat}$, using nonperiodic boundary conditions, by diagonalization within a complete basis of
$P(P+1)/2$ singlet states and $3P(P-1)/2$ triplet states, which yields the exact ground-state energy $E_{\rm gs}$ and
the exact ground-state density and magnetization, $n_k$ and $\bfm_k$, on each lattice site.

The extended Hubbard model described above can be treated using SDFT; \cite{Capelle2013} the corresponding Kohn-Sham lattice Hamiltonian
follows from $\hat H_{\rm lat}$ by setting $U_0=U_1=0$ and replacing
$V_k$ and $\bfB_k $ by the Kohn-Sham scalar potential and magnetic field $V_{{\rm KS},k}$ and $\bfB_{{\rm KS},k}$, respectively.
We invert the lattice analog of the Kohn-Sham equation, Eq. (\ref{2}), to obtain those $V_{{\rm KS},k}$ and $\bfB_{{\rm KS},k}$ which
reproduce the exact $n_k$ and $\bfm_k$ on the $P$-point lattice. Since the external potential and magnetic field are given,
and the Hartree potential is easily obtained as
\begin{equation}\label{VH}
V_{{\rm H},k}= U_0 n_k + U_1(n_{k-1}+n_{k+1})
\end{equation}
(defining $n_0=n_{P+1}=0$ to account for the nonperiodic boundary conditions), this immediately yields the
exact $V_{{\rm xc},k}$ and $\bfB_{{\rm xc},k}$. There exist a variety of
algorithms to invert Kohn-Sham equations; \cite{Jensen2017} here, we use a conjugate-gradient minimization (see Supplemental Material \cite{supp}).

Our choice to include n.n. interactions in $\hat H_{\rm lat}$ is motivated by the fact that for the usual o.s. Hubbard model (where $U_1=0$)
the exact exchange potential and exchange magnetic field are given by \cite{Ullrich2018,Ullrich2019,Carrascal2015}
\begin{equation}\label{os}
V_{{\rm x},k}^{\rm o.s.}=-\frac{U_0}{2}n_k \:, \qquad
\bfB_{{\rm x},k}^{\rm o.s.}=-\frac{U_0}{2} \bfm_k  \:.
\end{equation}
Evidently, $\bfB_{{\rm x},l}^{\rm o.s.}$ is parallel to the magnetization; hence, for the o.s. Hubbard model, the xc torques are a pure correlation effect
which, one can expect, makes them more difficult to approximate. For exchange effects to produce torques, more general interactions than o.s. are required.

Let us first consider the half-filled Hubbard dimer ($P=2$): \cite{Ullrich2018} including n.n. interactions, one finds
\begin{equation} \label{nn}
V_{{\rm x},k}^{\rm dimer}= -U_1-\frac{U_0-U_1}{2}n_k \:, \quad
\bfB_{{\rm x},k}^{\rm dimer}=-\frac{U_0-U_1}{2} \bfm_k \:,
\end{equation}
which, again, does not produce any torques. We therefore need more lattice points.

For the $1/3$-filled Hubbard trimer \cite{Ullrich2019} ($P=3$) with periodic boundary conditions and n.n. interactions, i.e., an equilateral triangle, Eq. (\ref{nn}) holds on all three lattice points.
If one instead considers a Hubbard trimer with fixed boundaries (i.e., a short chain), one finds that the solution (\ref{nn}) holds on the midpoint, but
there is no simple explicit solution on the first and the third point. Thus, exchange torques can arise on points 1 and 3, but not on point 2.

\begin{figure}[t]
\includegraphics[width=\linewidth]{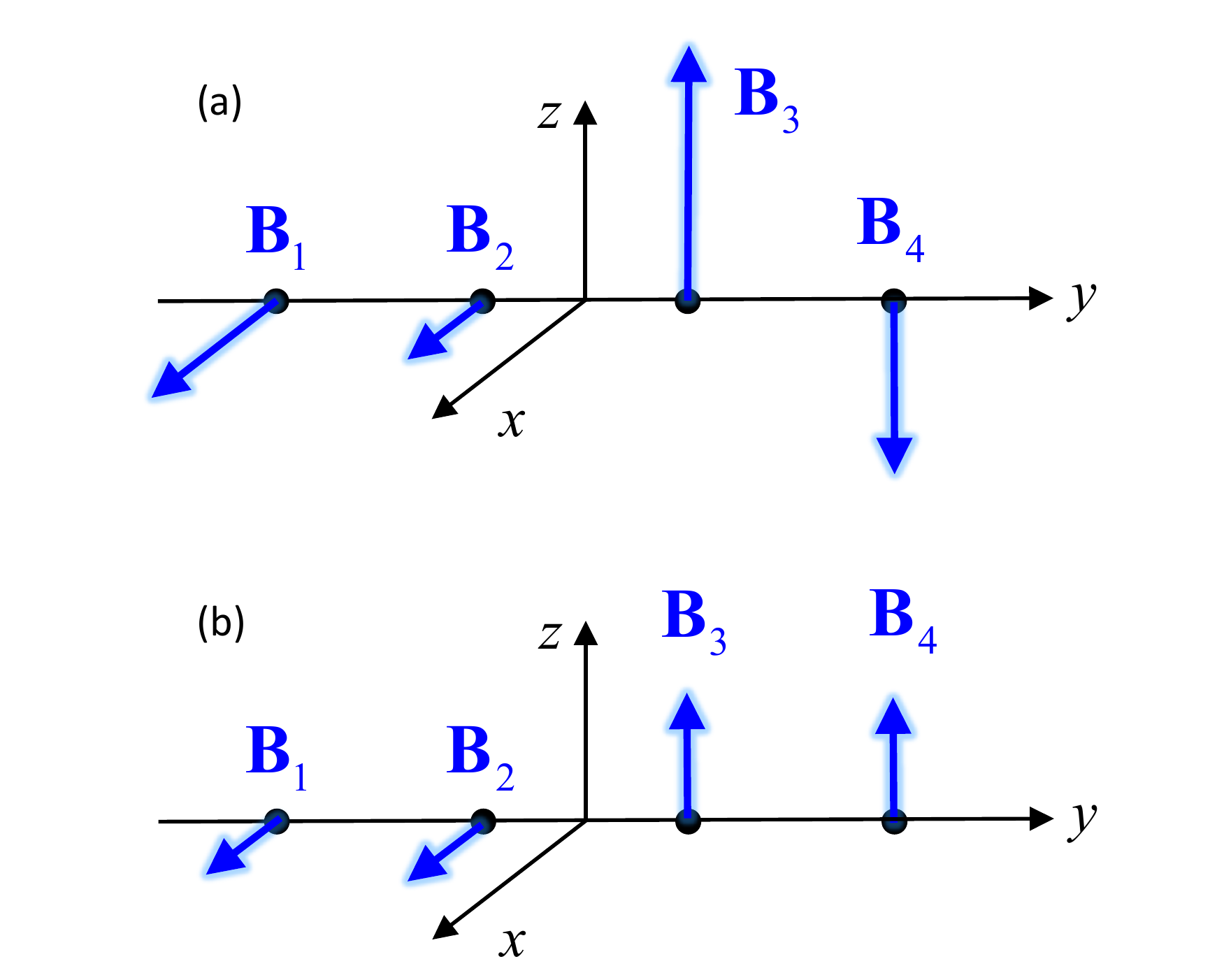}
\caption{(Color online)
Hubbard tetramer with noncollinear coplanar magnetic field $\bfB$ (blue arrows).
(a) Nonsymmetric field: $B_{x,1}=0.2$, $B_{x,2}=0.1$, $B_{z,3}=0.3$, $B_{z,4}=-0.2$.
(b) $C_2$-symmetric field: $B_{x,1} = B_{x,2} = B_{z,3} = B_{z,4} = 0.1$.
In both cases, the scalar potential is $V_1 = V_4 = 1$, $V_2 = V_3 = -1$. All local external and xc torques are along the $\pm y$-direction.
} \label{fig1}
\end{figure}

The simplest case where exchange torques can occur on all lattice sites is the $1/4$-filled Hubbard tetramer ($P=4$) with o.s. and n.n. interactions.
In the following, we will consider tetramers in linear configuration, see Fig. \ref{fig1}.
Of course, correlation effects can produce torques in all the above systems, even without n.n. interactions.

A two-electron system without external magnetic field is always in a singlet ground state and hence nonmagnetic.
To obtain a ground state with noncollinear magnetization, external magnetic fields are required (in the absence of spin-orbit coupling. \cite{Kaplan1983,Tabrizi2019})

In the following Sections, two cases with different symmetries will be considered.
In both of them, the 4-point lattice is assumed to have equidistant sites, aligned along the $y$-direction
(notice that the Hubbard lattice Hamiltonian is not necessarily tied to a specific real-space representation; however, assuming an explicit spatial geometry is conceptually
helpful). The scalar potential will be taken to be nonuniform but symmetric with respect to the lattice center; we choose $V_1 = V_4 = 1$ and $V_2 = V_3 = -1$
in both cases.

We restrict the magnetic fields and the resulting magnetization to be coplanar, confined to
the $x-z$ plane on each lattice point, but overall noncollinear (see Fig. \ref{fig1}). The resulting xc torques will lie along the $\pm y$-direction.

We obtain the ground-state energy, density, and magnetization on the lattice by exact diagonalization and via self-consistent Kohn-Sham.
For the latter, we use three approximations: \cite{Ullrich2018} exact exchange using the optimized effective potential approach (XX),
exact exchange within the Slater approximation (XXS), and xc using the Singwi-Tosi-Land-Sj\"{o}lander (STLS) scheme. \cite{Singwi1968,GiulianiVignale}
The latter is implemented within the scalar approximation, as detailed in Ref. \onlinecite{Ullrich2018}.

XX is fully variational and hence satisfies the zero-torque theorem; XXS and STLS, on the other hand, are not variational and can violate the zero-torque theorem.
As shown in the Supplemental Material, \cite{supp} the zero-torque theorem can be enforced by constrained optimization.
The torque-corrected approximations defined in this way will be denoted as XXSc and STLSc, respectively.

\begin{figure}
\includegraphics[width=\linewidth]{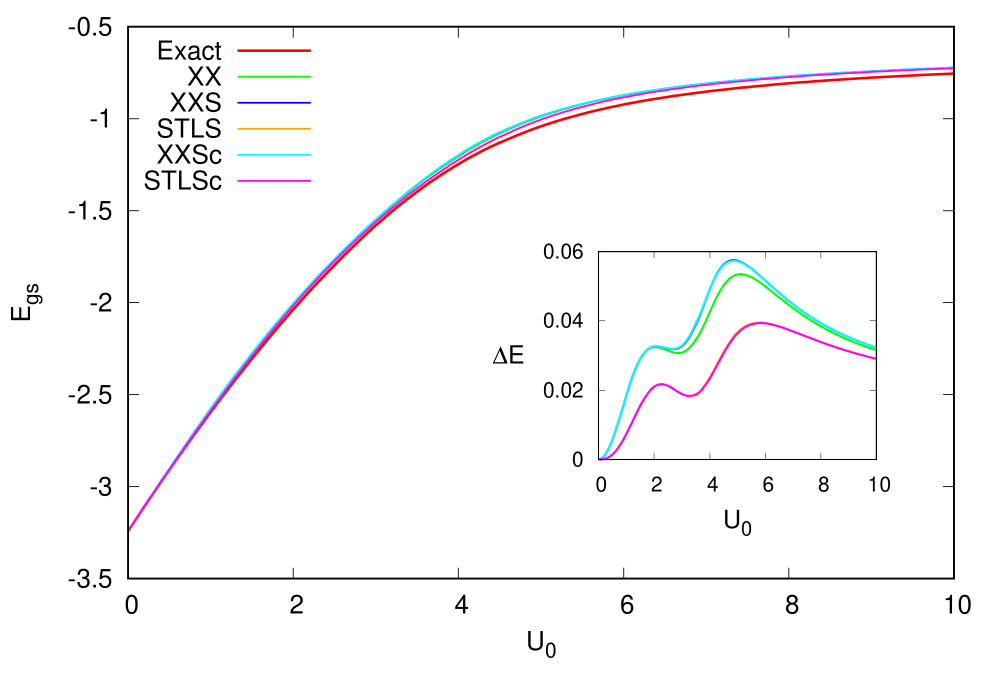}
\caption{(Color online) Ground-state energy of the Hubbard tetramer with nonsymmetric magnetic fields, see Fig. \ref{fig1}a,
comparing the exact result with various SDFT approximations (see text).
The difference between the exact solution and the SCDF approximations are shown in the inset.
The lines for XXS and XXSc, as well as STLS and STLSc, respectively, are too close to be distinguished.} \label{energy-asymmetric}
\end{figure}

\section{Nonsymmetric lattice} \label{sec:III}

We begin our discussion by considering a Hubbard tetramer where no distinct spatial or spin symmetries
are present. In this nonsymmetric case (see Fig. \ref{fig1}a), we set $B_{x,1} = 0.2$, $B_{x,2} = 0.1$, $B_{z,3} = 0.3$, and $B_{z,4} = -0.2$, with all other components zero.
The external magnetic field defined in this way is coplanar in the $x-z$ plane, and hence all torques (external as well as xc) are along the (positive or negative)
$y$-direction.

\subsection{Ground state energy}

In Fig. \ref{energy-asymmetric} we compare the exact ground-state energy of the nonsymmetric lattice with Kohn-Sham results using XX, XXS, STLS, XXSc and STLSc.
The exact energy starts out at $E_{\rm gs}=-3.242$ for the noninteracting case ($U_0=0$), reaches the value $E_{\rm gs}=-0.754$ for
$U_0=10$,  and approaches a limiting value of $E_{\rm gs}=-0.6107$ for $U_0 \to \infty$. The crossover between the weakly and strongly correlated
regimes occurs around $U_0 \sim 4$.
The agreement between the exact energy and the Kohn-Sham energies is very good for all values of $U_0$.
The inset shows that the largest errors of $E_{\rm gs}$  occur around $U_0=5$, reaching
around 5\% for XX and XXS and slightly less for STLS.

The xc torque correction of XXS and STLS has a negligible impact on the total energy; in Fig. \ref{energy-asymmetric}, the lines for XXS and XXSc are
essentially on top of each other, as are the lines for STLS and STLSc. The reason for this is easy to see:
to lowest order perturbation theory, the energy shift caused by a magnetic field $\bfB'(\bfr)$ is
\begin{equation}
\Delta E = \int d\bfr \: \bfm(\bfr) \cdot \bfB'(\bfr)  \:,
\end{equation}
where $\bfm(\bfr)$ is the magnetization of the unperturbed ground state. Thus, a transverse magnetic field [which is perpendicular to $\bfm(\bfr)$ for all $\bfr$]
only contributes to the total energy to second and higher order. The xc torque correction, see Eq. (30) of the Supplemental Material, \cite{supp} has the form of
a transverse magnetic field added to the xc magnetic field to be corrected.

\begin{figure}
\includegraphics[width=.99\linewidth]{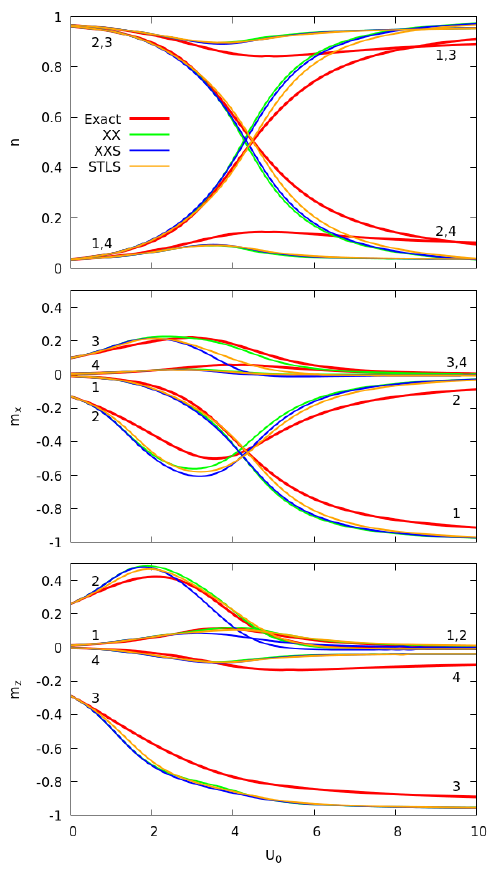}
  \caption{(Color online) Density (top) and $x$- and $z$-components (middle and bottom) of the magnetization of the nonsymmetric Hubbard tetramer.
  The numbers indicate lattice sites.} \label{n-asymmetric}
\end{figure}

\begin{table}
\caption{Exact density and magnetization data for the nonsymmetric and $C_2$-symmetric lattices, in the noninteracting $(U_0=0)$ and
strongly correlated $(U_0=10)$ limit.}
\begin{tabular}{c|cccc|cccc}
\hline \hline
 & \multicolumn{4}{c}{nonsymmetric} & \multicolumn{4}{c}{$C_2$-symmetric} \\
$U_0$  &  $n_1$ & $n_2$ & $n_3$ & $n_4$ &  $n_1$ & $n_2$ & $n_3$ & $n_4$\\ \hline
0  & 0.037 & 0.961 & 0.967 & 0.034 & 0.036 & 0.964 & 0.964 & 0.036\\
10 & 0.911 & 0.095 & 0.891 & 0.102 & 0.518 & 0.482 & 0.482 & 0.518\\[3mm]
&  $m_{1}$ & $m_{2}$ & $m_{3}$ & $m_{4}$ &  $m_{1}$ & $m_{2}$ & $m_{3}$ & $m_{4}$ \\ \hline
0  & 0.017 & 0.291 & 0.304 & 0.005 & 0.007 & 0.143 & 0.143 & 0.007 \\
10 & 0.911 & 0.088 & 0.890 & 0.102 & 0.518 & 0.477 & 0.477 & 0.518\\[3mm]
  &  $m_{x,1}$ & $m_{x,2}$ & $m_{x,3}$ & $m_{x,4}$ &  $m_{x,1}$ & $m_{x,2}$ & $m_{x,3}$ & $m_{x,4}$\\ \hline
0  & -0.010 & -0.131 & 0.099 & 0.005 & -0.005 & -0.103 & 0.098 & 0.005\\
10 & -0.911 & -0.088 & 0.005 & 0.004 & -0.518 & -0.477 & 0.004 & 0.009\\[3mm]
  &  $m_{z,1}$ & $m_{z,2}$ & $m_{z,3}$ & $m_{z,4}$  &  $m_{z,1}$ & $m_{z,2}$ & $m_{z,3}$ & $m_{z,4}$\\ \hline
0  &  0.013 &  0.260 & -0.288 & -0.002 &  0.005 &  0.098 & -0.103 & -0.005\\
10 & 0.007 & -0.006 & -0.890 & -0.102 & 0.009 & 0.004 & -0.477 & -0.518\\\hline \hline
\end{tabular}
\label{table1}
\end{table}

\subsection{Density and magnetization}

Figure \ref{n-asymmetric} shows the density $n_k$ (top panel) and the $x$- and $z$-components of the magnetization, $m_{x,k}$ and $m_{z,k}$ (middle and bottom panels).
At $U_0=0$, the density is almost entirely concentrated on points 2 and 3 (where the scalar potential is lower). As $U_0$ increases,
the electronic repulsion leads to a redistribution of the site occupation, so the electrons can better avoid each other (keep in mind that
our model includes the n.n. interaction $U_1$); the crossover clearly occurs around $U_0 \sim 4$.
At $U_0=10$, lattice sites 1 and 3 are almost fully occupied,
whereas sites 2 and 4 are almost empty, see Table \ref{table1} for details.
This behavior is somewhat exaggerated, but overall well reproduced, by the approximate Kohn-Sham calculations.
Again, XXSc and STLSc give results very similar to XXS and STLS, and are not shown here.

The magnetization reveals further details which illustrate the behavior of the system as the interaction strength increases.
At $U_0=0$, the exact two-electron wave function has 95\% singlet character, \cite{note}
and the system is only weakly magnetized (see Table \ref{table1}).
On the other hand, at $U_0=10$, the exact wave function has 74\% triplet character, and the magnitude of the magnetization
vector on each lattice site is practically identical with the density, which indicates a fully magnetized state.
In the limit $U_0\to \infty$, the total magnetization adds up to 2, and the singlet-triplet ratio is 1:3.

Furthermore, Table \ref{table1} tells us that at $U_0=10$ the magnetization vector is
(anti)aligned with the applied magnetic field on each lattice site: on sites 1 and 2, the magnetization
points almost completely in the negative $x$-direction, and on sites 3 and 4 it points in the negative $z$-direction.
Referring back to Fig. \ref{fig1}a, we see that the magnetization vector is opposite to the applied
magnetic field on sites 1, 2, and 3, and parallel to it on site 4.
The physical meaning of this is that in the strongly correlated limit, the electrons  not only localize spatially,
they also align their spin quantization axis with the local magnetic field.

\begin{figure}
\includegraphics[width=.9\linewidth]{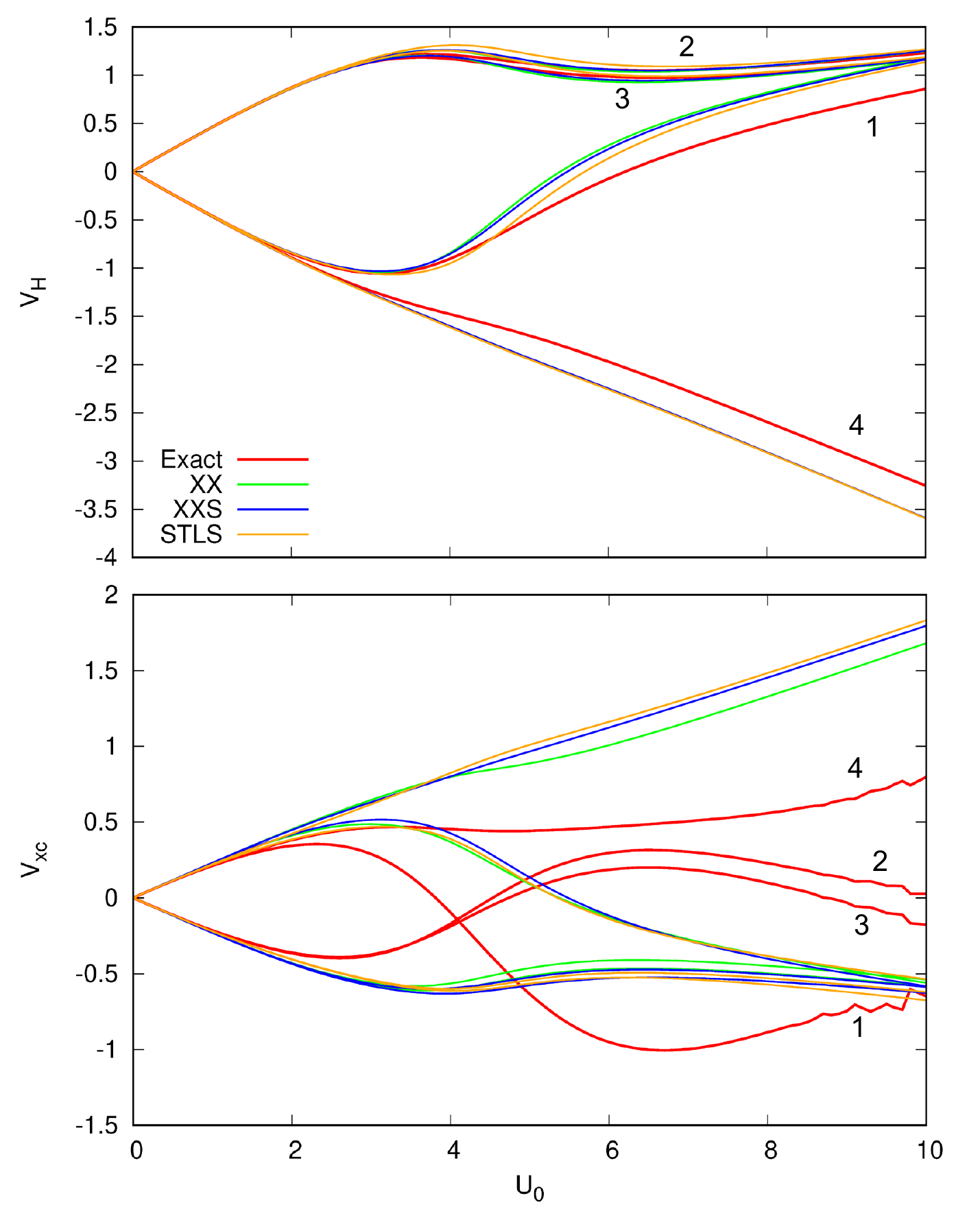}
  \caption{(Color online) Hartree (top) and xc potential (bottom) of the nonsymmetric Hubbard tetramer.
  The numbers indicate lattice sites.} \label{VHXC-asymmetric}
\end{figure}

\begin{figure}
\includegraphics[width=.9\linewidth]{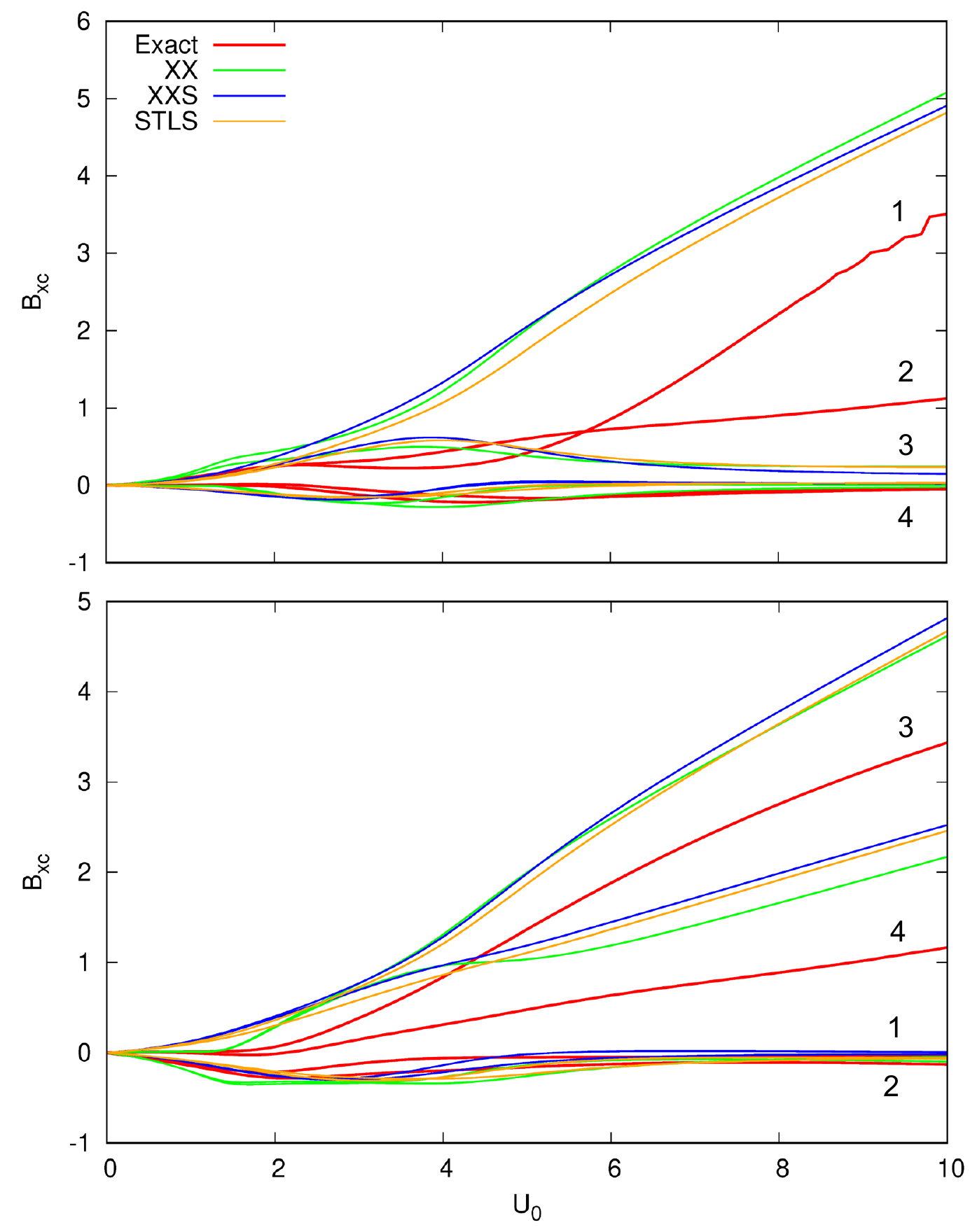}
  \caption{(Color online) $x$- (top) and $z$-component (bottom) of the xc magnetic field of the nonsymmetric Hubbard tetramer.
  The numbers indicate lattice sites.} \label{BXC-asymmetric}
\end{figure}

\subsection{Hartree and xc potentials and xc magnetic fields}
Figure \ref{VHXC-asymmetric} compares the exact Hartree potential, Eq. (\ref{VH}), and the exact xc potential, respectively, with the corresponding
XX, XXS(c) and STLS(c) approximations. The $x$- and $z$-components of the xc magnetic fields are shown in Fig. \ref{BXC-asymmetric}.

$V_{{\rm H},k}$, $V_{{\rm xc},k}$, and $\bfB_{{\rm xc},k}$ are responsible for reproducing the exact density and magnetization
in the noninteracting Kohn-Sham system. As the system passes from the weakly to the strongly interacting regime with increasing $U_0$,
density gets shifted from site 2 to site 1, and sites 1 and 3 become almost fully magnetized. The Kohn-Sham system accomplishes this
through an intricate interplay between Hartree and xc effects. For instance, comparing $V_{{\rm H},k}$ and $V_{{\rm xc},k}$, we see that the xc potential
partially counteracts the Hartree potential, especially on sites 1 and  4.

The xc magnetic field displays a striking enhancement over the external magnetic field, by about an order of magnitude. These large field strengths
are necessary so that the Kohn-Sham system can create the alignment of the local magnetic moments with the external field in the strongly correlated limit.
Looking again at Table \ref{table1}, we see that the magnetic moments are almost completely aligned along the $x$-direction on sites 1 and 2,
and along the $z$-directions on sites 3 and 4. Correspondingly, the $x$- and $z$-components of $\bfB_{{\rm xc},k}$ are very strongly enhanced on these sites.

The approximate xc potentials and magnetic fields agree well with the exact results until about $U_0=3$. For larger interaction strengths, considerable
differences appear. For example, the xc potentials on sites 1, 2, and 3 change sign, which is not reproduced by any of the approximations,
and the xc potential on site 4 has the wrong limiting behavior for large $U_0$. Similar discrepancies are observed for the xc magnetic fields.
This helps explain the differences between exact and approximate densities and magnetizations in Fig. \ref{n-asymmetric}.

\begin{figure}
  \includegraphics[width=.99\linewidth]{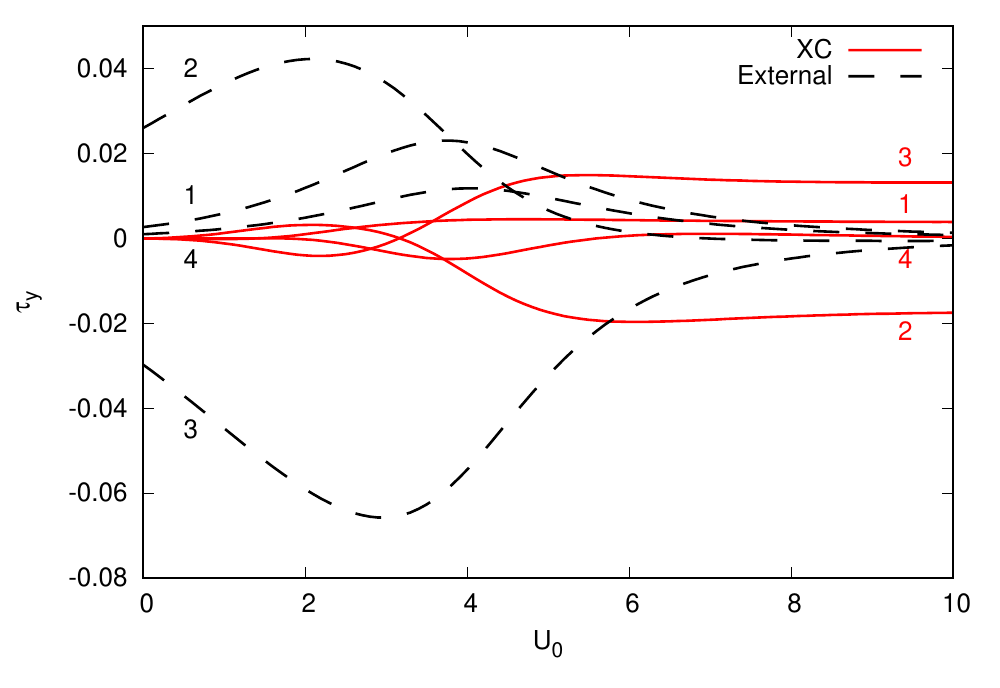}
  \caption{(Color online) Comparison of xc torque  $\bm{\tau}_{{\rm xc},k}$ (red, full lines) and external torque  $\bm{\tau}_{{\rm ext},k}$ (black, dashed lines)
  of the nonsymmetric tetramer. All torques only have a $y$-component.} \label{tmag-asymmetric}
\end{figure}

\begin{figure*}
  \includegraphics[width=.99\linewidth]{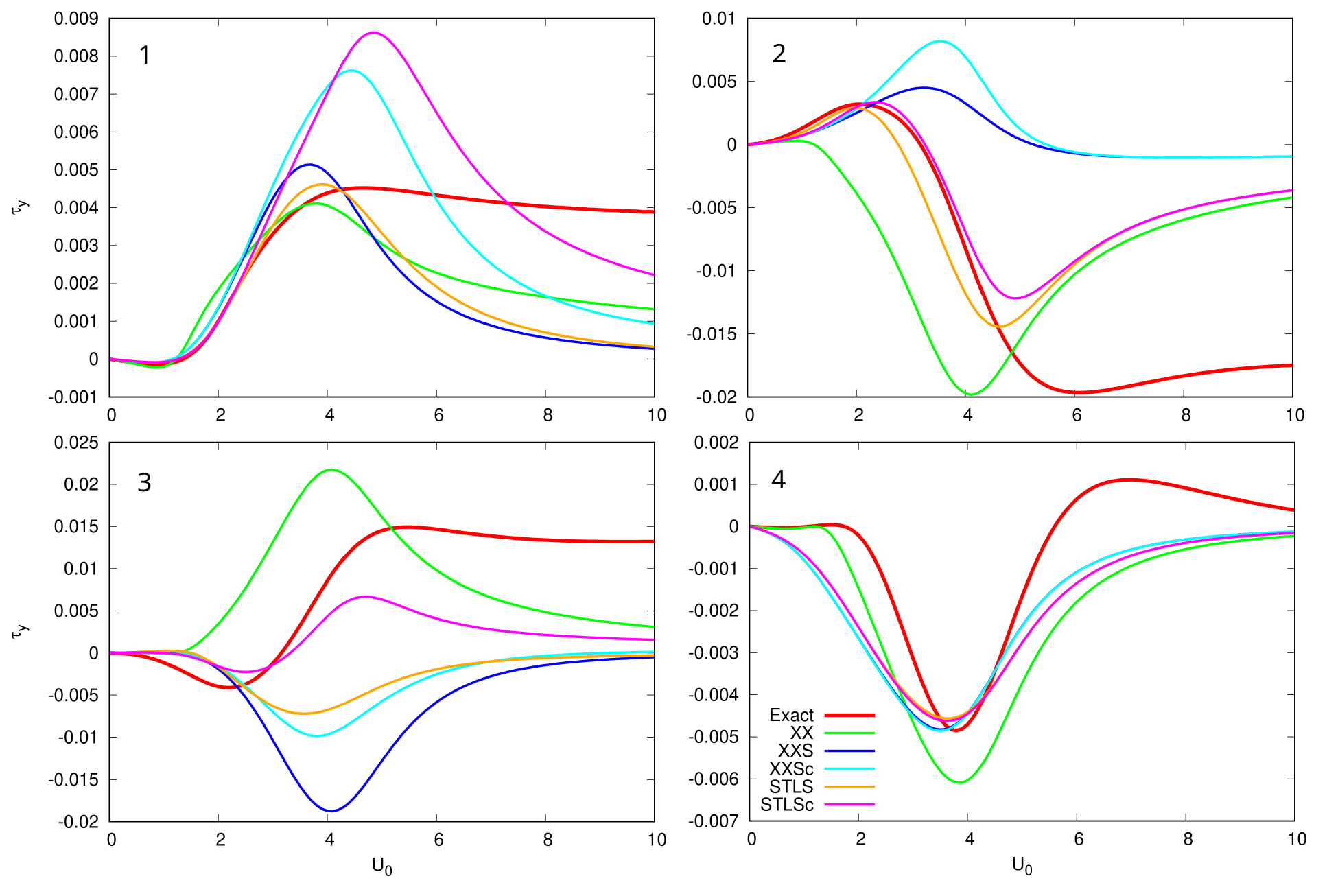}
  \caption{(Colors online) Comparison of exact xc torque  $\bm{\tau}_{{\rm xc},k}$ and xc torque approximations with and without torque correction for the nonsymmetric tetramer,
  on the four lattice sites. All torques only have a $y$-component.} \label{ty-asymmetric}
\end{figure*}

\subsection{Exact and approximate xc torques}

Let us now examine the xc torques of the nonsymmetric tetramer. Figure \ref{tmag-asymmetric} shows a comparison of the exact $\bm{\tau}_{{\rm xc},k}$ (full lines)
with the torque caused by the externally applied magnetic field, $\bm{\tau}_{{\rm ext},k}$ (dashed lines). The latter is defined in analogy with Eq. (\ref{tau}),
with $\bfB_{{\rm xc},k}$ replaced by $\bfB_k$. Both torques are along the $y$-direction on all lattice sites. It can be seen that
the xc torques on each lattice site add up to zero, as do the external torques (as they must, since the system is in equilibrium).

For weak to moderate interaction strengths, the external torque dominates over the xc torque, with maxima around $U_0=2$ to 4, depending on the lattice site.
As the system localizes for $U_0 \gtrsim 4$, the magnetization on each lattice site aligns with the external magnetic field,
as discussed in the previous subsection, so that $\bm{\tau}_{{\rm ext},k}$
goes to zero.

The xc torques, on the other hand, are small for weak interactions, but becomes larger than the external torque on sites 2 and 3 after the crossover into the strongly correlated
regime. This means that, on these sites, the xc magnetic field does not quite align with the local magnetization. Looking at Table \ref{table1},
we see that at $U_0=10$  sites 2 and 3 are almost, but not fully magnetized and aligned with the local external fields; furthermore, as Fig. \ref{BXC-asymmetric}
shows, the xc magnetic fields are very large on sites 2 and 3. Together, this results in a finite $\bm{\tau}_{{\rm xc},k}$, which is necessary for the Kohn-Sham
system to reproduce all subtle details of the magnetization of the interacting system.

The xc torques obtained with XX, XXS, STLS, XXSc and STLSc are presented in Fig. \ref{ty-asymmetric} for each lattice point.
The xc torques are much smaller on sites 1 and 4 than on sites 2 and 3, which means that the performance
of the approximations is more crucial on the latter sites. Compared to the exact xc torques,
the approximate xc torques generally have the right order of magnitude, which is reassuring; however, there are some important details, depending on the
lattice site and the level of approximation.

On site 1, all approximations perform well for small to moderate interactions, but fail to reproduce the finite xc torque for large interaction strengths;
XXSc and STLSc overshoot around $U_0=5$. On site 4, the quantitative agreement is quite good throughout, for all approximations.

On sites 2 and 3, the exact xc torque changes sign between $U_0=3$ and 4. This trend is not reproduced by XX, XXS and XXSc.
STLS performs much better for small to moderate interaction strengths, including the sign change, with STLSc leading to slight improvements over STLS.
However, all approximations produce a vanishing xc torque in the strongly correlated regime, instead of approaching a finite limit.

How important are the xc torques? We have already seen that they give only a very small contribution to the total energy.
We have carried out self-consistent calculations with XX, XXS and STLS where we only include the longitudinal components of $\bfB_{{\rm xc},k}$
(see the Supplemental Material \cite{supp} for details). We find that the magnetization on the individual sites does not change much for the
nonsymmetric case.

\section{Lattice with $C_2$ symmetry} \label{sec:IV}

We now consider a magnetic field distribution which has $C_2$ symmetry, as illustrated in Fig. \ref{fig1}b. The symmetry is with
respect to rotation about an axis which lies at an angle of 45$^{\rm o}$ between the $x$ and $z$ axes. Specifically, we choose
$B_{x,1} = B_{x,2} = B_{z,3} = B_{z,4} = 0.1$. In the Supplemental Material \cite{supp} we show that the resulting magnetization
also has $C_2$ symmetry. Specifically, one finds $m_{x,k} = m_{z,P-k+1}$, and a similar relation for the xc magnetic field.
Hence, the zero-torque theorem will automatically be satisfied for all approximations (as long as the Kohn-Sham solution maintains
the symmetry).

\begin{figure}[t]
  \includegraphics[width=\linewidth]{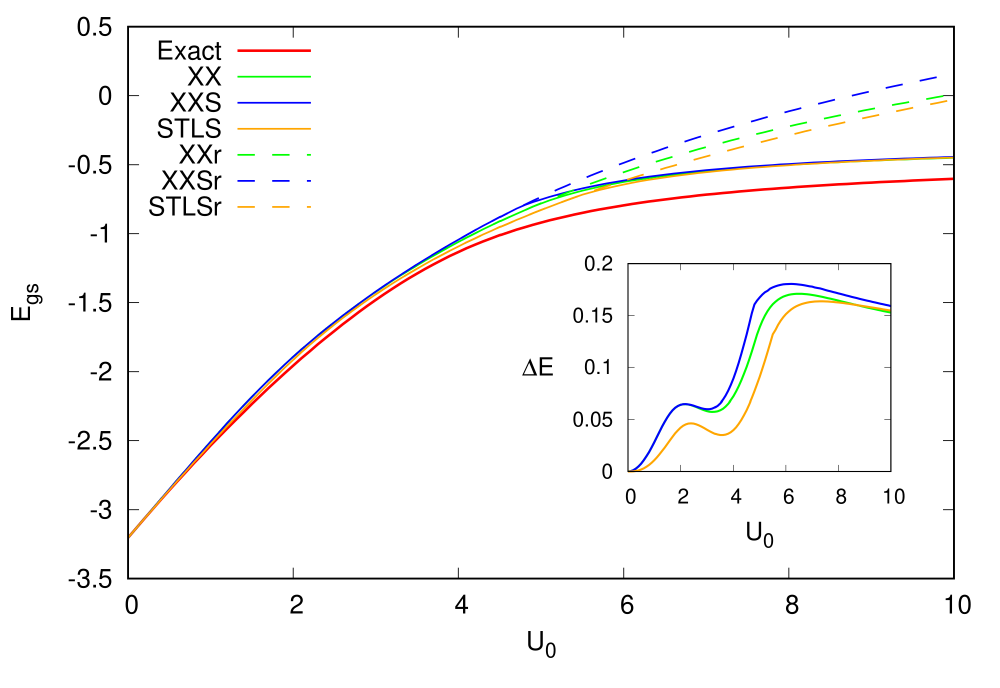}
  \caption{(Color online) Ground-state energy of the Hubbard tetramer with $C_2$ symmetry. Dashed lines are symmetry enforced calculations, while the solid lines were allowed to break the symmetry of the system. The difference between the exact solution and the corresponding (unrestricted) approximations are plotted in the inset.} \label{energy-symmetric}
\end{figure}

\begin{figure}[t]
  \includegraphics[width=\linewidth]{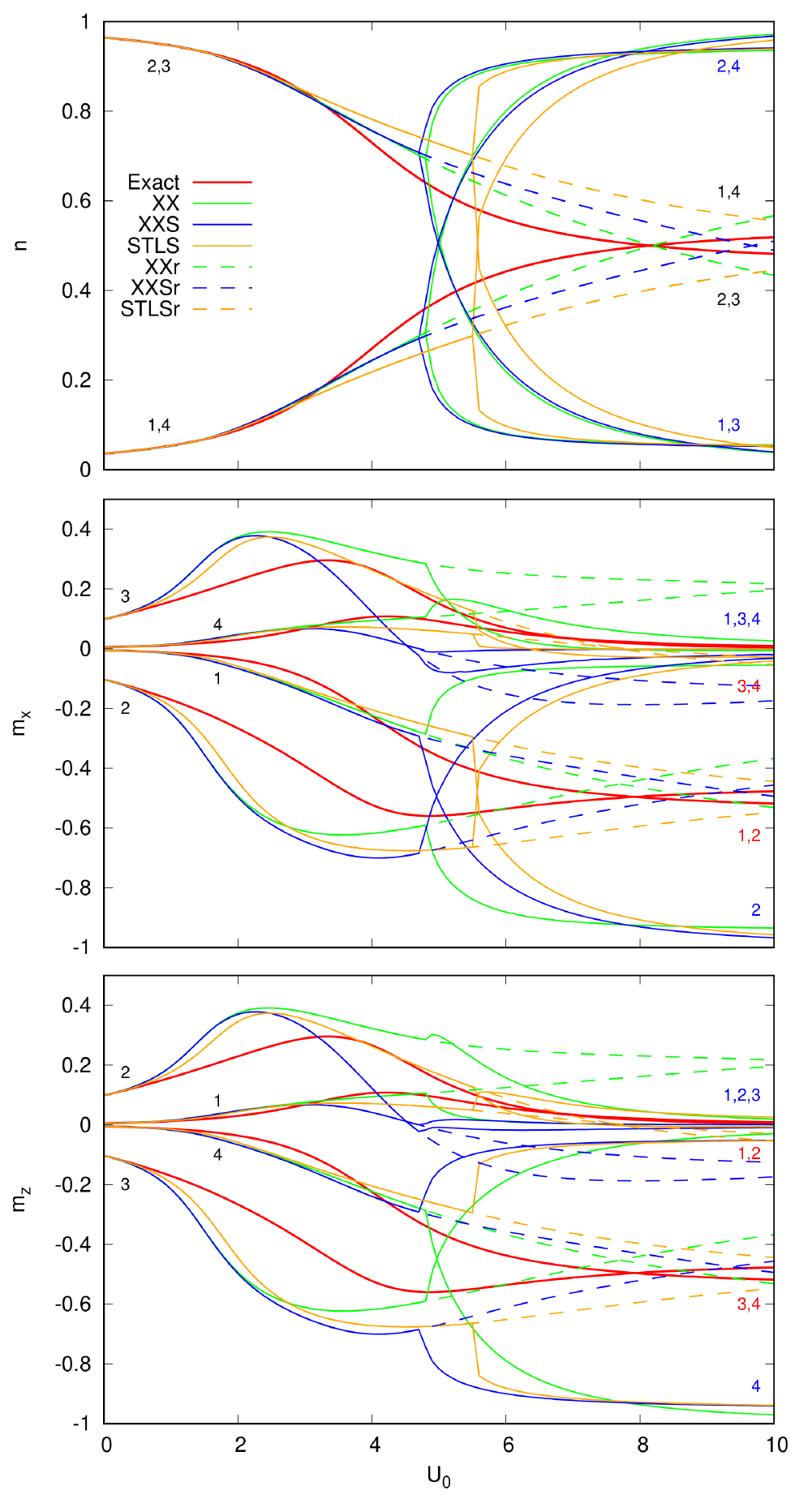}
   \caption{(Color online) Density (top) and $x$- and $z$-components (middle and bottom) of the magnetization of the Hubbard tetramer with $C_2$ symmetry.
  The numbers indicate lattice sites.} \label{n-symmetric}
\end{figure}

\begin{figure}
\includegraphics[width=.9\linewidth]{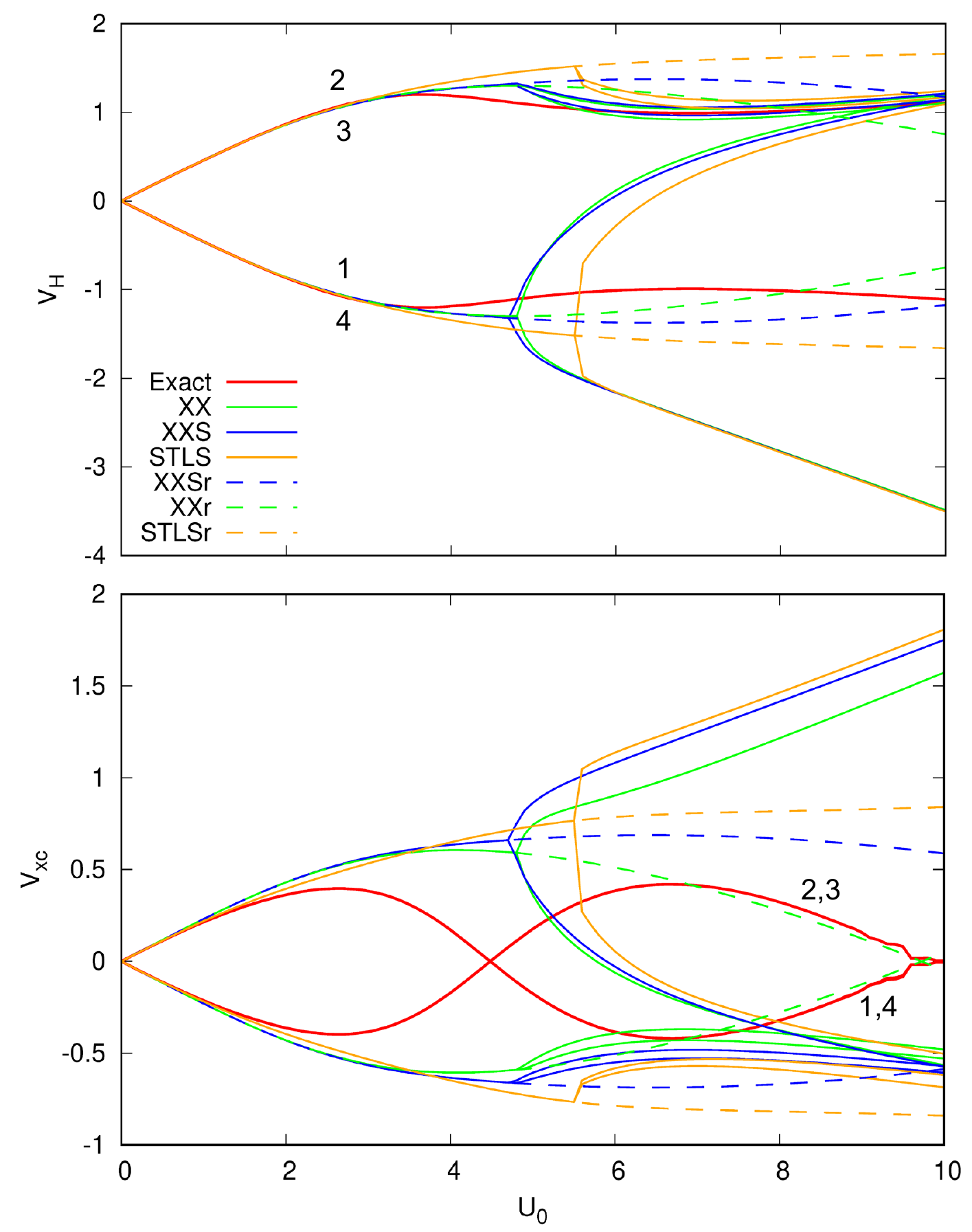}
  \caption{(Color online) Hartree (top) and xc potential (bottom) of the $C_2$-symmetric Hubbard tetramer.
  The numbers indicate lattice sites.} \label{VHXC-symmetric}
\end{figure}

\begin{figure}
\includegraphics[width=.9\linewidth]{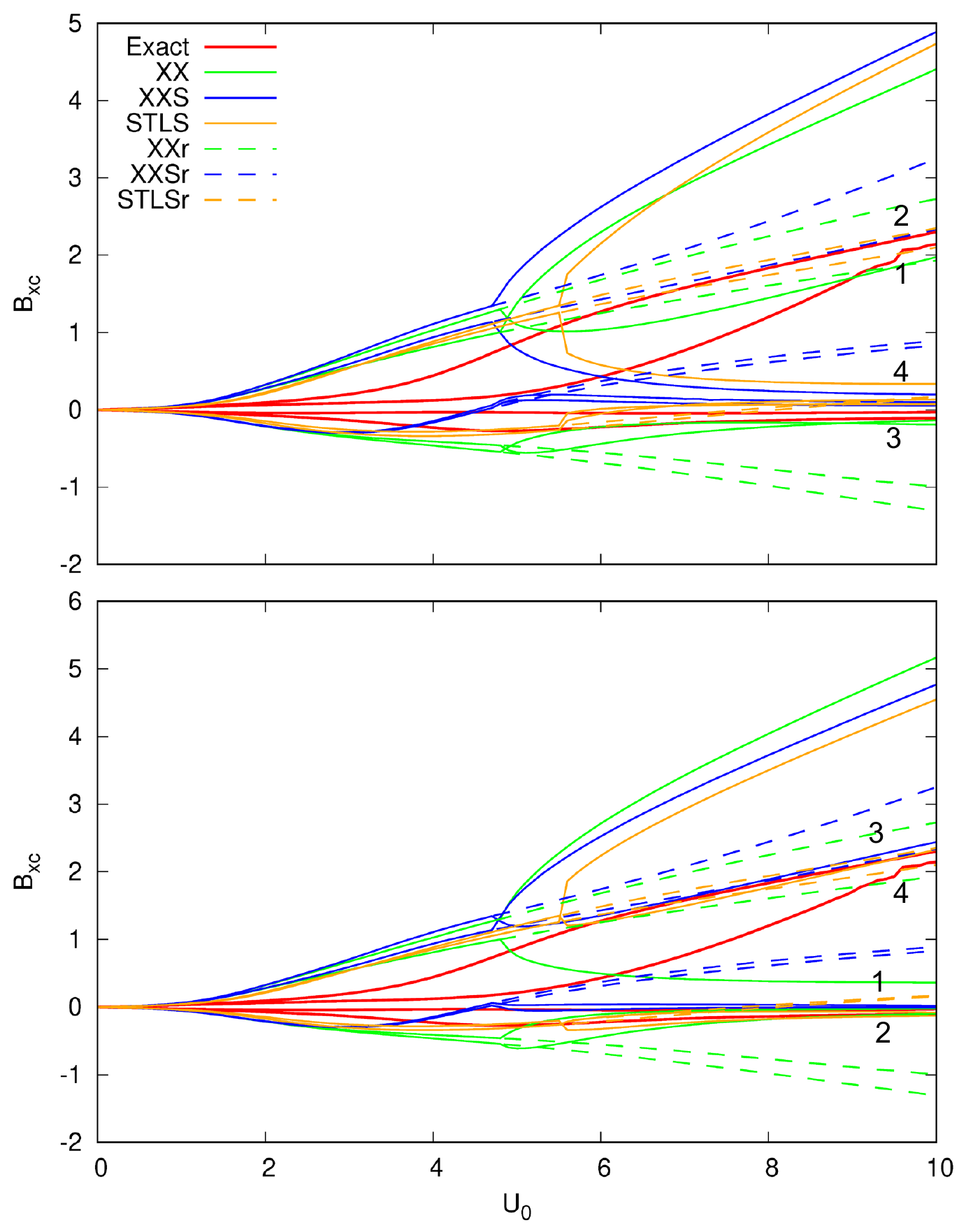}
  \caption{(Color online) $x$- (top) and $z$-component (bottom) of the xc magnetic field of the $C_2$-symmetric Hubbard tetramer.
  The numbers indicate lattice sites.} \label{BXC-symmetric}
\end{figure}

\begin{figure}
  \includegraphics[width=\linewidth]{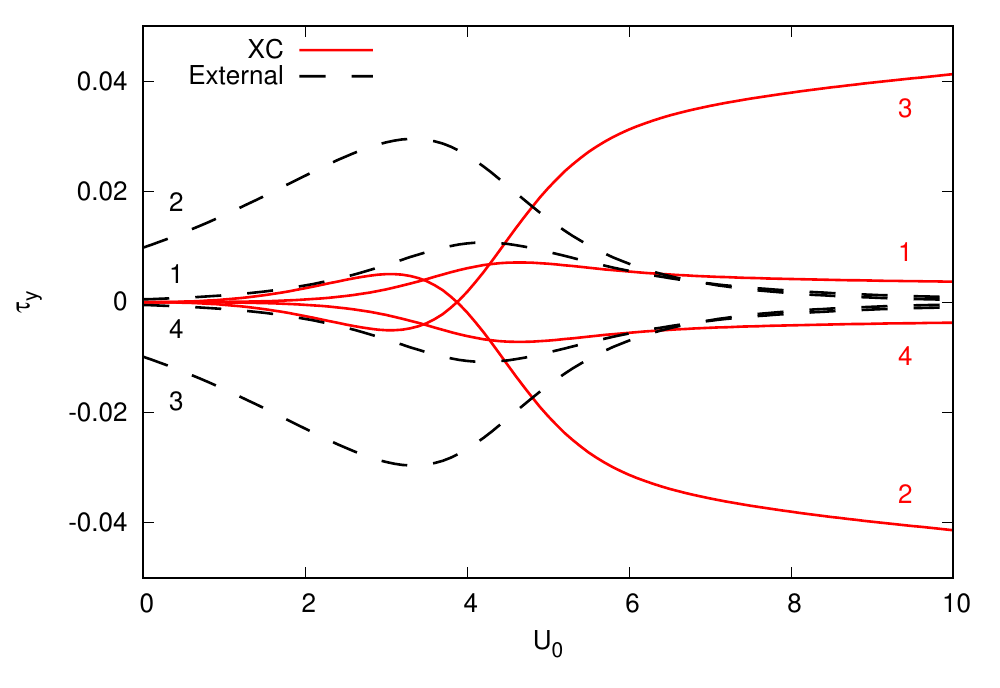}
  \caption{(Color online) Comparison of xc torque  $\bm{\tau}_{{\rm xc},k}$ (red, full lines) and external torque  $\bm{\tau}_{{\rm ext},k}$ (black, dashed lines)
  of the $C_2$-symmetric tetramer.   All torques only have a $y$-component.} \label{tmag-symmetric}
\end{figure}

\begin{figure*}
  \includegraphics[width=.99\linewidth]{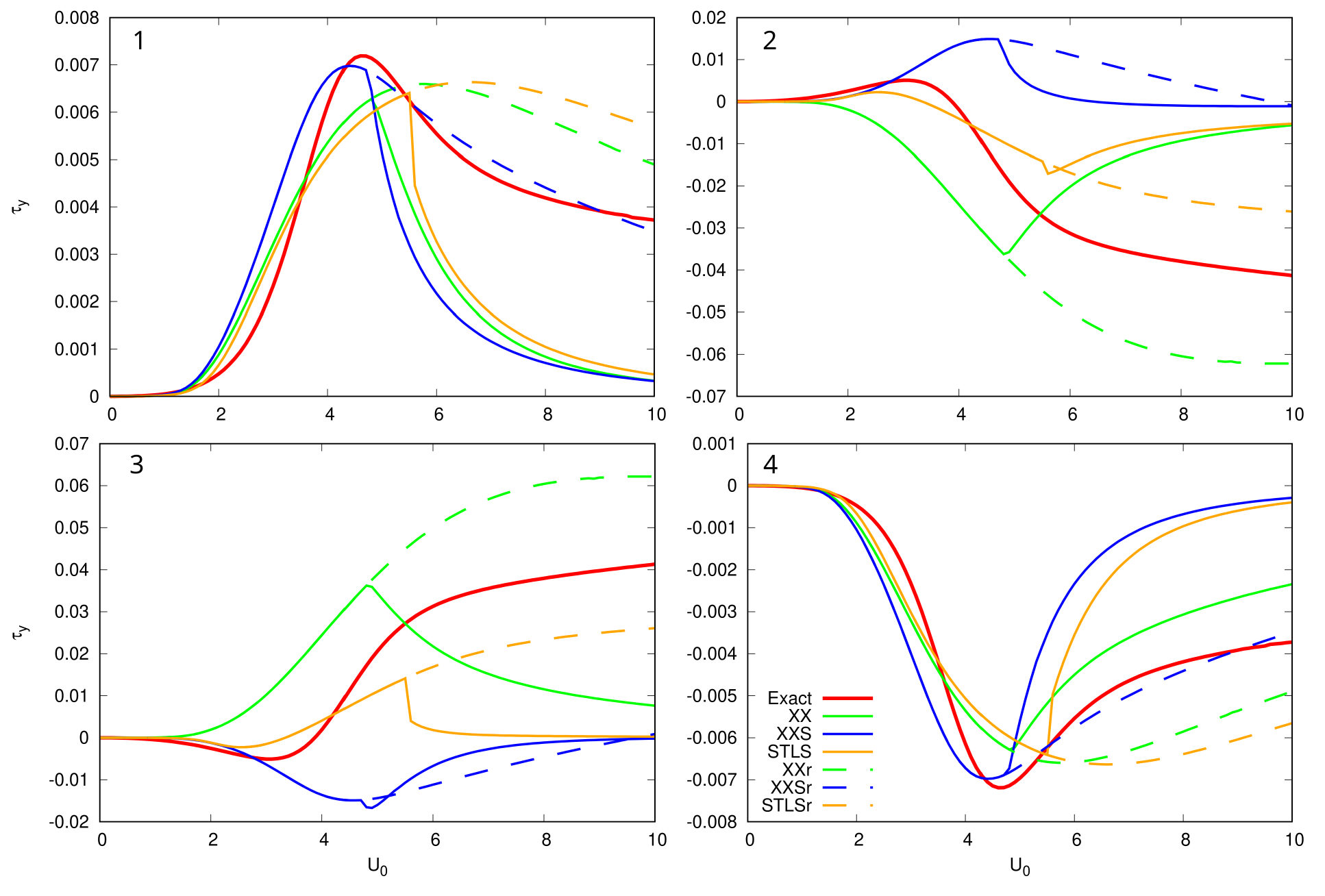}
  \caption{(Colors online) Comparison of exact xc torque  $\bm{\tau}_{{\rm xc},k}$ and unrestricted and restricted xc torque approximations for the $C_2$-symmetric tetramer,
  on the four lattice sites.  All torques only have a $y$-component.} \label{ty-symm}
\end{figure*}

\subsection{Ground-state energy, density and magnetization: breaking and enforcing symmetry}

Figure \ref{energy-symmetric} shows the ground-state energy $E_{\rm gs}$ for the $C_2$-symmetric Hubbard tetramer for interaction
strengths $U_0$ between 0 and 10. The behavior is similar to the nonsymmetric case discussed earlier.
The exact energy starts out at $E_{\rm gs}=-3.203$ for the noninteracting case, reaches the value $E_{\rm gs}=-0.603$ for
$U_0=10$,  and approaches a limiting value of $E_{\rm gs}=-0.4247$ for $U_0 \to \infty$. As before, the crossover between the weakly and strongly correlated
regimes occurs around $U_0 \sim 4$.

The agreement between the exact energy and the (unrestricted) SDFT results is generally good for weak and moderate interactions strengths,
but a significant ($\sim$25\%) deviation appears for large $U_0$.
The inset to Fig. \ref{energy-symmetric} shows the difference between approximate and exact ground-state energies. Overall, STLS performs
the best; in the limit of large $U_0$, XX merges with STLS.

Each approximation reaches a critical $U_0$ where it is advantageous for the system to break $C_2$ symmetry to reduce the total energy of the system.
This does not come as a surprise: \cite{Jacob2012,Carrascal2015,Ullrich2018} unrestricted Kohn-Sham calculations with approximate xc functionals have the general tendency to
break symmetries (spatial and/or spin) to lower the energy. This is known as the symmetry dilemma of DFT.
The dashed lines in Fig. \ref{energy-symmetric} are the restricted SDFT ground-state energies, obtained by enforcing the $C_2$ symmetry.
The restricted energies keep increasing with $U_0$, in contrast with the exact $E_{\rm gs}$ and with the unrestricted (symmetry-broken) Kohn-Sham $E_{\rm gs}$.

Figure \ref{n-symmetric} shows the density $n_k$ (top panel) and the $x$- and $z$-components of the magnetization, $m_{x,k}$ and $m_{z,k}$ (middle and bottom panels).
At $U_0=0$, the density is almost entirely concentrated at points 2 and 3, like in the nonsymmetric case (see the top panel of Fig. \ref{n-asymmetric}). However,
as $U_0$ increases, differences between the symmetric and nonsymmetric tetramers start to appear: for strong correlations,
the exact density becomes almost evenly distributed over the four lattice points, rather than accumulating on two points.
After breaking the symmetry, the unrestricted approximate SDFT calculations dramatically over-localize the density in the strongly correlated limit: the density
is almost completely concentrated on points 2 and 4. The symmetry breaking occurs at $U_0=4.72$ and $U_0=4.78$ for XXS and XX, respectively, and
at $U_0=5.24$ for STLS.

For the exact magnetization (middle and bottom panels of Fig. \ref{n-symmetric}, and right half of Table \ref{table1}),
the trends are very similar to the nonsymmetric case. At $U_0=0$, the system is only very weakly magnetized, with an
exact wave function that has 99\% singlet character. At $U_0=10$, the system is fully magnetized, and the exact wave function has 74\% triplet character.
As before, the exact magnetization vector is (anti)aligned with the applied magnetic field on each lattice site, which indicates
a local spin quantization axis.

Unrestricted SDFT, on the other hand, produces a very different magnetization after the $C_2$ symmetry is broken.
As we discussed above, the unrestricted density localizes on points 2 and 4, and we see that $m_{x,2}$ and $m_{z,4}$ approach $-1$.
As for the exact case, the total magnetization adds up to 2 in the $U_0\to \infty$ limit, in spite of the wrong overall distribution.

The dashed lines in Fig. \ref{n-symmetric} show the density and magnetization resulting from symmetry-restricted SDFT.
Not surprisingly, the agreement with the exact results is much better, although there are still some deviations.
However, the price to pay is a wrong behavior of the total energy, as we saw above in Fig. \ref{energy-symmetric}.

\subsection{xc potentials, magnetic fields, and torques}
Figure \ref{VHXC-symmetric} shows the exact Hartree potential and the exact xc potential on each lattice site, as a function of $U_0$.
The $x$- and $z$-components of the xc magnetic fields are shown in Fig. \ref{BXC-symmetric}. The interpretation of these results is
a bit more straightforward than for the nonsymmetric case discussed earlier (see Figs. \ref{VHXC-asymmetric} and  \ref{BXC-asymmetric}).
The Hartree potential approaches values close to 1 on sites 2 and 3 and close to $-1$ on sites 1 and 4, which compensates the external
potential. The xc potential is small compared to the Hartree potential (the fact that it approaches zero for $U_0=10$ seems accidental).
Together, the Kohn-Sham potential is close to uniform, which explains the almost uniform density distribution in the strongly
interacting limit, see Table \ref{table1}. The strong xc magnetic fields, on the other hand, are responsible for almost fully magnetizing the
system, similar to what we discussed earlier for the nonsymmetric case.

As before for the nonsymmetric case, the XX, XXS and STLS approximations for $V_{{\rm xc},k}$ and $\bfB_{{\rm xc},k}$ deviate significantly
from the exact results, especially after the symmetry breaking. The agreement within the restricted-symmetry formalism is somewhat better.

Figure \ref{tmag-symmetric} shows a comparison of the exact $\bm{\tau}_{{\rm xc},k}$ (full lines) for the $C_2$-symmetric tetramer
with the torques caused by the externally applied magnetic field, $\bm{\tau}_{{\rm ext},k}$ (dashed lines). As in the nonsymmetric case (Fig. \ref{tmag-asymmetric}),
the external torque dominates over the xc torque for weak interactions. However, in the strongly correlated regime, the xc torques are now
considerably larger (in magnitude as well as relative to the external torques) than in the nonsymmetric case.

The xc torques obtained with XX(r), XXS(r) and STLS(r) are presented in Fig. \ref{ty-symm}.
For small to moderate $U_0$ the agreement with the exact xc torques is decent, especially on sites 1 and 4. The deviations on site 2 and 3
are more significant, however. As expected, the approximate xc torques fail completely for the unrestricted calculations after
symmetry breaking. The restricted calculations (dashed lines) perform somewhat better in general.
The excellent agreement between XXSr and the exact results on sites 1 and 4 seems, however, fortuitous.
STLS is most successful at capturing the general trends in both the weakly and strongly correlated regimes.

In the Supplemental Material \cite{supp} we show the xc torques for the $C_2$-symmetric lattice calculated using XXS and STLS with
the torque correction. Some minor differences can be observed, but the torque correction has no impact on the symmetry breaking.

Furthermore, in the Supplemental Material \cite{supp} we show results for the $C_2$-symmetric lattice calculated using only the
longitudinal components of $\bfB_{{\rm xc},k}$. As for the nonsymmetric lattice, the differences are only relatively minor, except
for XX: the XX functional seems to be more sensitive to the presence of the transverse xc magnetic field, which affects the way
in which the symmetry is broken. Overall, the longitudinal-only XX is closer to its XXS and STLS counterparts (and to the exact solution)
than when the full $\bfB_{{\rm xc},k}$ is used, which would suggest that XX does not describe transverse xc effects very well.
This is consistent with the behavior of the xc torques shown in Fig. \ref{ty-symm} (and, to some extent, also in Fig. \ref{ty-asymmetric}),
which shows that the XX torques tend to be somewhat exaggerated. The behavior of XXS and STLS comes across as more robust.

\section{Conclusions}\label{sec:V}

In this paper we have carried out a detailed computational study of a simple model system: two interacting electrons on a four-site extended
Hubbard lattice, in the presence of noncollinear magnetic fields. Our goal was to find out how a Kohn-Sham system, through its
xc scalar potentials and magnetic fields, manages to reproduce the exact ground-state density and magnetization of the interacting system, particularly
for situations where the interaction is strong. We discovered that the xc magnetic field plays a key role in ensuring that
the magnetic behavior of strongly correlated, localized electrons is determined by local spin quantization axes on each lattice site.

We compared the exact xc potentials and magnetic fields with several orbital-dependent approximations, and found that their performance
is overall quite good, but depends on whether the system has any symmetries. In the case where the Hubbard lattice has $C_2$-symmetry,
we found that the approximate Kohn-Sham calculations tend to break this symmetry once the interaction strength becomes sufficiently large,
thus lowering the energy.
This, of course, is no surprise: the so-called symmetry dilemma is a well-known side effect of using approximate xc functionals in SDFT.

A main outcome of this study is that the xc torques seem to have a relatively minor importance for the ground-state energy, density
and magnetization, compared to the xc potentials and longitudinal xc magnetic fields --- at least for the simple model systems considered here.
This suggests that the widely used standard approximations of noncollinear magnetism (which have no transverse magnetic xc contributions)
should be adequate in most situations, as long as they reproduce the longitudinal xc magnetic fields sufficiently accurately.

However, our study did not include spin-orbit coupling, which is an important aspect of noncollinear magnetism in many materials.
In particular, we needed to enforce the noncollinear magnetic state via applied magnetic fields, whereas
effects such as canted or frustrated spins arise naturally without any applied fields. It is an open question whether xc torque effects
play a more important role in such situations, but it seems not unlikely. To address this issue, Hubbard models that
include spin-orbit coupling are a promising approach.

A perhaps even more important question concerns the role of xc magnetic fields for spin dynamics. While the xc torques appear less crucial in the ground
state, they may become more important as the magnetic system evolves in time, following a sudden switching or a pulsed excitation.
Such studies, again based on simple Hubbard-type systems, are currently in progress.

Lastly, while simple lattice models serve as important benchmarks for SDFT and can produce a wealth of new physical insight,
it will be important to investigate the performance of orbital-dependent xc functionals for noncollinear magnetic systems in real space.
The XXS approximation will be very well suited for this purpose due to its simplicity and reasonable accuracy. The STLS approach
was found to perform well for the model systems considered here (as long as correlations are not too strong), but would be numerically much more expensive for real-space systems.
Finding orbital-dependent correlation functionals for noncollinear magnetic materials, at a reasonable computational cost, remains an ongoing challenge.

\acknowledgments

This work was supported by DOE Grant No. DE-SC0019109 and by a Research Corporation for Science Advancement Cottrell Scholar SEED Award.

\bibliography{XCTorque_refs}

\begin{thebibliography}{37}
\bibitem [{not()}]{note}%
  \BibitemOpen
  \href@noop {} {}\bibinfo {note} {The singlet/triplet character of the exact
  wave function follows from the exact diagonalization, which is carried out in
  a basis of noninteracting singlet and triplet states.}\BibitemShut {Stop}%
\end{thebibliography}

\end{document}